\pdfoutput=1
\RequirePackage{xcolor}
\makeatletter
\@ifundefined{set@color}{\def\set@color{}}{}
\makeatother

\documentclass{PoS}
\usepackage{epsfig}
\usepackage{cite}
\newcommand{\comment}[1]{}
\usepackage{lipsum}
\usepackage{enumitem}
\usepackage{wrapfig}
\usepackage{marvosym}
\usepackage{graphicx}
\usepackage{epsfig}
\usepackage{bm}
\usepackage{latexsym,amssymb,amsmath,amsfonts,amssymb,txfonts,pxfonts,wasysym,float}

\DeclareBoldMathCommand\blpar{\left(}
\DeclareBoldMathCommand\brpar{\right)}

\newcommand{\beq}[1]{\begin{equation}\label{#1}}
\newcommand{\eeq}{\end{equation}}
\newcommand{\bea}[1]{\begin{eqnarray} \label{#1}}
\newcommand{\eea}{\end{eqnarray}}
\newcommand{\ba}{\begin{array}}
\newcommand{\ea}{\end{array}}

\def\be{\begin{equation}}
\def\ee{\end{equation}}
\def\gs{\mathrel{
   \rlap{\raise 0.511ex \hbox{$>$}}{\lower 0.511ex \hbox{$\sim$}}}}
\def\ls{\mathrel{
   \rlap{\raise 0.511ex \hbox{$<$}}{\lower 0.511ex \hbox{$\sim$}}}}

\newcommand{\postscript}[2]{\setlength{\epsfxsize}{#2\hsize}
   \centerline{\epsfbox{#1}}}

\definecolor{orange}{cmyk}{0,0.5,1,0}
\definecolor{rossoCP3}{cmyk}{0,.88,.77,.40}
\definecolor{graa}{rgb}{0.8,0.8,0.8}
\definecolor{blaa}{rgb}{0.2,0.2,0.6}

\title{EUSO-SPB2 sensitivity to macroscopic dark matter}

\ShortTitle{EUSO-SPB2 sensitivity to macroscopic dark matter}

\author{\speaker{Thomas C. Paul}$^{1}$, Sarah T. Reese$^{2}$,
  Luis A. Anchordoqui$^{1,2,3}$, and Angela V. Olinto$^{4}$\\
  $^1$Department of Physics, Lehman College, City University of New
  York, NY 10468, USA\\
  $^2$Department of Physics, Graduate Center, City University
  of New York,  NY 10016, USA\\
$^3$Department of Astrophysics, American Museum of Natural History, NY
10024, USA\\
$^4$Department of Astronomy \& Astrophysics, University of Chicago, Chicago, IL 60637, USA
}

\abstract{\noindent Macroscopic dark matter (or macro) provides a broad class of alternative
candidates to particle dark matter. These candidates would transfer
energy primarily through elastic scattering, and this linear energy
deposition would produce observable signals if a macro were to traverse
the atmosphere. We study the fluorescence emission produced by a macro
passing through the atmosphere. We estimate the sensitivity of EUSO-SPB2
to constrain the two-dimensional parameter space ($\sigma$ vs. $M$),
where $M$ is the macro mass and $\sigma$ its cross sectional area.}

\FullConference{37th International Cosmic Ray Conference - ICRC2021 -\\
		15 - 22 July, 2021\\
	  Berlin, Germany}
\begin{document}

\section{Introduction}

The underlying goal of the particle physics program is to
discover the connection between the Standard Model (SM) and dark
matter (DM). For the last few decades, the favored DM model has been a
relic density of weakly interacting massive particles (or WIMPs)~\cite{Lee:1977ua,Vysotsky:1977pe,Goldberg:1983nd,Steigman:1984ac}. However, LHC experiments have run extensive physics searches
for WIMP fingerprints that have returned only null results~\cite{Penning:2017tmb,Rappoccio:2018qxp,Buchmueller:2017qhf}. There has also been a broad WIMP search program using direct and
indirect detection methods, which so far has also given unsatisfactory
answers~\cite{Undagoitia:2015gya,Gaskins:2016cha}. Even though a thorough exploration of the WIMP parameter space remains the highest priority of the DM community, there is now a strong motivation to seek alternatives to the WIMP paradigm.

In this communication we reexamine the well-motivated hypothesis which
postulates that rather than being intrinsically weakly interacting, DM can be
effectively weakly interacting because it is massive and hence has a much lower
number density. The rate of DM-baryon interactions go as $\sim n_{_{\rm DM}}
\ \sigma \ v$, where $n_{_ {\rm DM}} \sim \rho_{_{\rm DM}}/ M$ is the DM number
density, $\sigma$ is the DM-baryon scattering cross-section, and $v \sim
250~{\rm km/s}$ is the characteristic velocity of the Sun's galactic
rotation. The event rate is proportional to the reduced cross section $\sigma/M$
for a given $\rho_{_{\rm DM}}$, which in our galactic neighborhood is
approximately 1 proton-mass for every 3 cubic centimeters, i.e. $\rho_{_{\rm
    DM}} \sim 7 \times 10^{-25}~{\rm g/cm^3}$~\cite{Zyla:2020zbs}. In the WIMP
paradigm DM has thus far evaded detection because the particle species has a
weak-scale mass and interaction strength; dimensional analysis gives $\sigma
\sim g^4/(4 \pi M)^2 \sim 10^{-8}~{\rm GeV}^{-2}$, with $M \sim G_F^{-1/2} \sim
245~{\rm GeV}$ and the coupling $g \sim 0.65$, yielding $\sigma/M \sim 4 \times
10^{-11}~{\rm GeV}^{-3}$. However, it can equally be that DM interacts strongly,
but has escaped detection because $M \gg M_{\rm Pl} \sim 10^{19}~{\rm GeV}$;
e.g., for a macroscopic DM particle (a.k.a. macro) of $M \sim 10^{18}~{\rm g}$
we would expect DM to hit the Earth about once every billion years. Indeed, the
local flux of DM is estimated to be ${\cal F}_{_{\rm DM}} \sim v \rho_{_{\rm
    DM}} \sim 1.7 \times 10^{-17}~{\rm g \ cm^{-2} \ s^{-1}}$ and thus the
annual Earth infall is roughly $10^9~{\rm g}$~\cite{DeRujula:1984axn}.

Among the most appealing macro candidates is Witten's
proposal~\cite{Witten:1984rs} wherein the QCD phase transition in the early
universe resulted in an abundance of baryons alongside macroscopically sized
nuggets of strange quark matter~\cite{Farhi:1984qu,Alcock:1988re}, with a
density $\rho_s \sim 3.6 \times 10^{14}~{\rm
  g/cm^3}$~\cite{Chin:1979yb}. A related proposal is that of dark
quark nuggets~\cite{Bai:2018dxf}. Another possibility for macroscopic DM is to be a
population of primordial black holes with $M \gtrsim 10^{15}~{\rm
  g}$~\cite{Chapline:1975ojl,Carr:2020xqk,Villanueva-Domingo:2021spv}, which are
unaffected by Hawking radiation~\cite{Hawking:1974rv}. While specific models
have their own charm (or better say strangeness), we find it sagacious to
reexamine the phenomenology of generic models in which DM interacts strongly,
with interaction probability determined predominantly by geometry and
kinematics. More concretely, we investigate the feasibility to search for macro
interactions in the Earth's atmosphere using the second generation Extreme
Universe Space Observatory on a Super-Pressure Balloon (EUSO-SPB2) mission,
which has been approved by NASA for a long duration flight in
2022~\cite{Wiencke:2019vke}.

The layout of the paper is as follows. We
begin in Sec.~\ref{sec:2} with a concise description of the EUSO-SPB2
mission. In Sec.~\ref{sec:3} we study the interactions produced by a
macro passing through the atmosphere and characterize the signal visible to the EUSO-SPB2
fluorescence detector. The paper wraps up
with some conclusions presented in Sec.~\ref{sec:4}.

\section{Generalities of the EUSO-SPB2 mission}

\label{sec:2}

\begin{wrapfigure}{r}{0.5\textwidth}
  \postscript{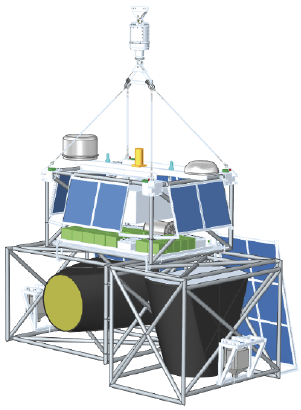}{0.6}
\caption{EUSO-SPB2 design with two telescopes: a fluorescence telescope and a Cherenkov telescope. The fluorescence telescope can observe at nadir, while the Cherenkov telescope is aimed at the limb of the Earth.}
\label{fig:1}
\end{wrapfigure}

The EUSO-SPB2 mission will monitor the night sky of the Southern hemisphere to
pioneer Earth-orbiting observation of cosmic rays of very high to ultrahigh
energies and to search for upward-going showers generated by energetic $\tau$
leptons produced by $\tau$-neutrino interactions in the Earth's limb.  The
payload, now in the design and fabrication stage, features a pair of telescopes:
one looking down on the atmosphere with an optical fluorescence detector from
the near space altitude of $h = 33~{\rm km}$, and the other looking towards the
limb of the Earth to observe Cherenkov signals; see Fig.~\ref{fig:1}. Both
telescopes are based on a Schmidt optical design of spherical mirrors with large
fields of view (FoVs). For a down pointing (nadir), the FoV opening angle is
$12^\circ$. The telescopes have a larger FoV for limb observations.

A launch from NASA's Wanaka facility is motivated by
the opportunity for a flight of up to 100 days. Wanaka lies under a
stratospheric air circulation that forms each Fall and Spring. During
the Austral Fall, this wide river of high thin air circles the
Southern globe to the east at 50 to 150~km/h. A March/April launch
window can insert the balloon into this circulation. The balloon and
telescopes with an apparent wind speed of essentially
zero. The super pressure balloon is sealed. At float it expands to a
maximum constrained volume with a slight over-pressure that remains
positive at night to maintain the fixed volume. Hence the balloon
floats at a nearly constant altitude. In the absence of technical
issues, EUSO-SPB2 can remain aloft almost indefinitely.

\begin{figure*}[ht]
  \postscript{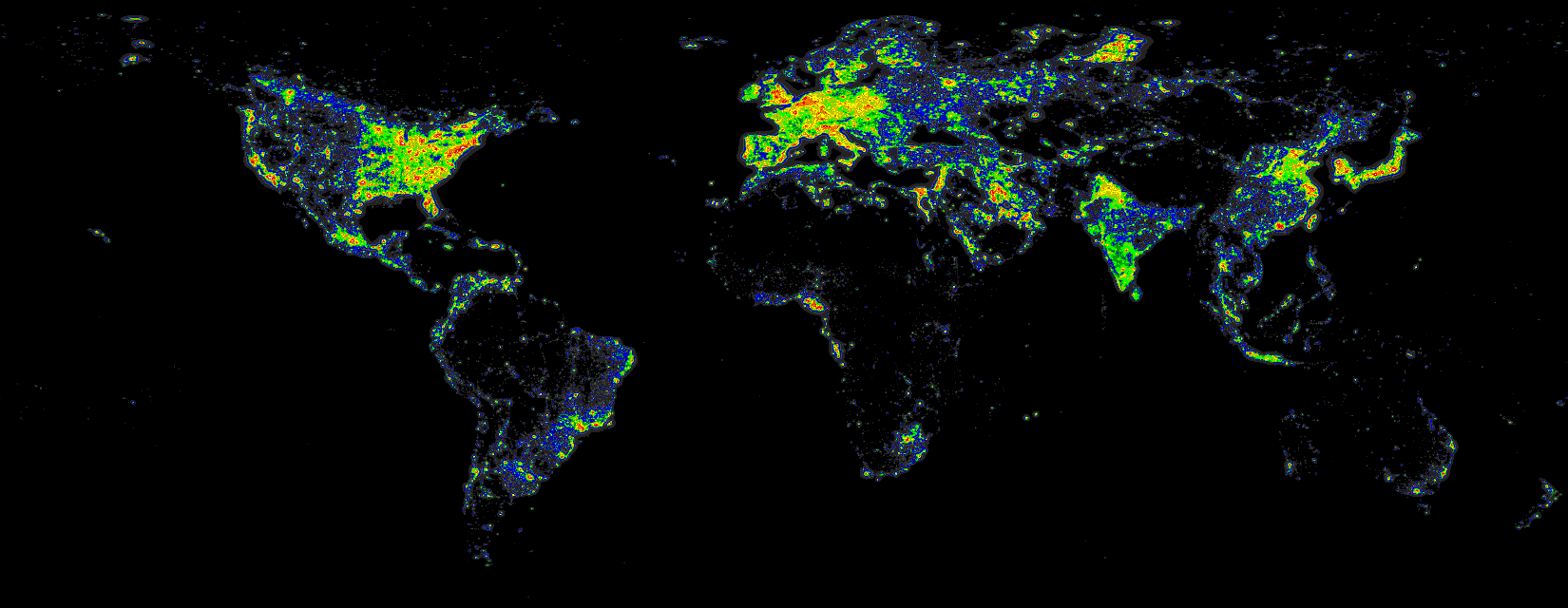}{1.0}
\caption{Light pollution heat map of the world.}
\label{fig:2}
\end{figure*}

Observations can only be done on clear moonless nights. As EUSO-SPB2
will fly in the Southern hemisphere it is subject to relatively little
pollution, see Fig.~\ref{fig:2}. A thorough study suggests that a 50
day flight launched at Wanaka (latitude of $45^\circ~{\rm S}$) during
the March/April window would see between 190 and 260~hr of dark time,
depending on when the launch happens relative to the moon
phase~\cite{Wiencke:2015oko}.  For a 100~day flight, the fluctuations
would smooth out a bit, and so about 500~hr would be a realistic
number of dark hours, with no moon and between the end and start of
astronomical twilight at 33~km. This estimate does not take into
account possible reduction of the duty cycle due to obscuration by
clouds, and further assumes an operationally perfect detector. When
trigger effects and reconstruction efficiency in the presence of
clouds are taken into account the effective observational time is
estimated to be $t_{\rm obs} \sim 360~{\rm hr}$~\cite{Adams:2017fjh}.

For a nadir pointing instrument that has a constant full FoV, the
instantaneous observational area (ignoring the
Earth curvature) is a circle on the ground of radius $r = h \tan({\rm FoV}/2)$. The instantaneous
observational area of the fluorescence telescope is then
\begin{equation}
A = \pi \ \bigg[h \tan \left(\frac{{\rm FoV}}{2} \right) \bigg]^2 = 32~{\rm km}^2 \, .
\end{equation}
When this area is combined with the effective observational time of
360~hr we obtain a total expected
exposure of ${\cal E} = 1.2 \times 10^4~{\rm km}^2 \, {\rm hr}$.

\section{Macro's fluorescence signal and EUSO-SPB2 sensitivity to the
  DM flux}
\label{sec:3}

Like meteoroids, macros are susceptible to rapid heat loss upon entering the  Earth's atmosphere as a result of elastic collisions
with the air molecules. Actually, it is at lower altitudes where the
macro encounters the exponentially increasing atmospheric density and
undergoes rapid heating along its path, which expands and
radiates. There are a tremendous number of mean free
paths/interactions involved as the macro bowling ball rolls over the
ping-pong air molecules. As it traverses the atmosphere the macro
transfers each particle it hits a velocity in the gas frame of ${\cal
  O} (v)$. The rate of energy loss per
unit path-length is given by the number of molecules of the atmosphere
scattered by the macro per unit path-length
times the average energy transfer per molecule hit~\cite{DeRujula:1984axn}:
\begin{equation}
  \frac{dE}{dx} = -
  \sigma \ \rho_{_{\rm atm}} \ v^2 \,,
\label{uno}
\end{equation}
where $\rho_{_{\rm atm}} (\zeta) \sim 10^{-3} \ e^{-\zeta/(8~{\rm km})}~{\rm
    g/cm}^3$ is the atmospheric density at altitude $\zeta$~\cite{Anchordoqui:2018qom}. We are interested though in the
fraction 
of dissipated energy  appearing as UV
light, $\eta$. Following~\cite{DeRujula:1984axn, Sidhu:2018auv}, we estimate a lower bound on the
luminosity efficiency from thermodynamical arguments in which light is
emitted from an expanding cylindrical shock wave. The
thermodynamic description of the event is valid if the macro's radius,
\begin{equation}
  r_{_0} =  8.7 \times 10^{-6} \ \left(\frac{M}{{\rm g}} \right)^{1/3} \
  \left(\frac{\rho_s}{\rho_m}\right)^{1/3}~{\rm cm} \,,
    \end{equation}
    exceeds the macro's mean-free-path in air, where $\rho_m$ is the macro density. 

\begin{figure}
\postscript{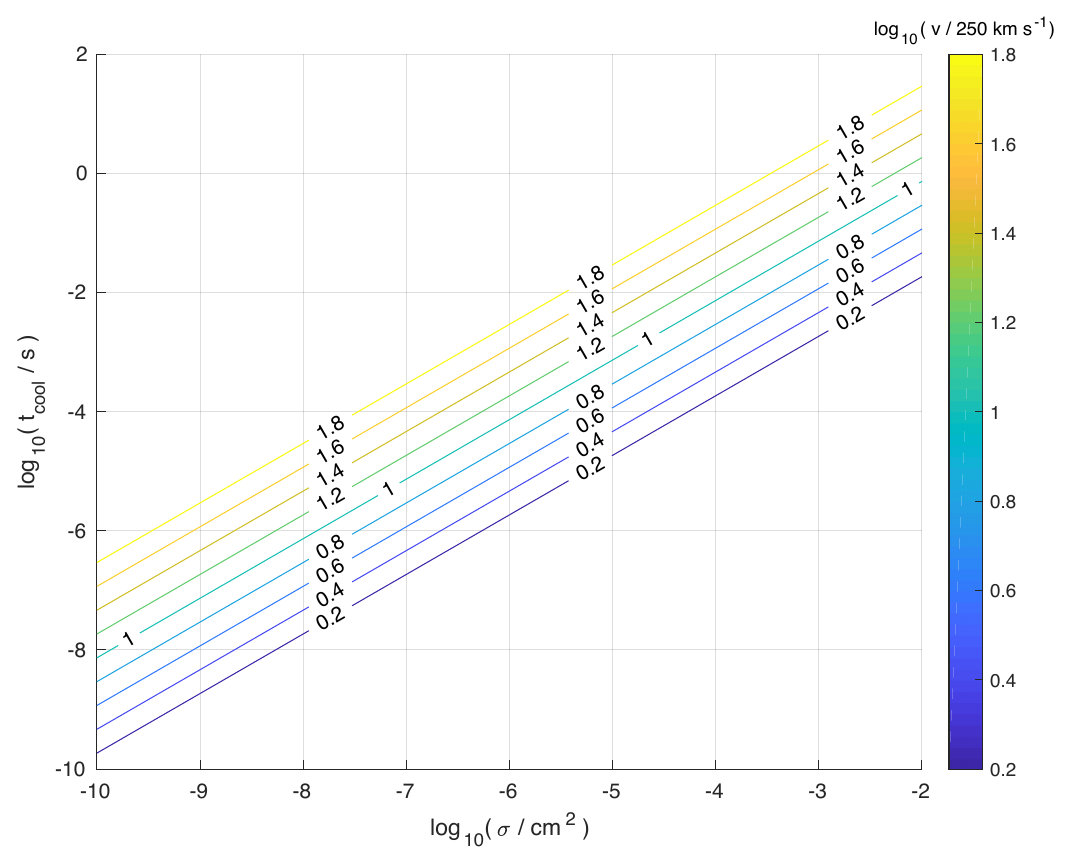}{0.6}
\caption{The cooling time as a function of the macro's cross-sectional area for
  a Maxwellian distribution of velocity.
  \label{fig:3}}
\end{figure}

A more restrictive condition emerges by demanding that the thermal fluctuations in the plasma,
    \begin{equation}
      \frac{\langle (E - \langle E \rangle )^2 \rangle}{\langle E
        \rangle^2 }\sim {\cal V}^{-1/2} \,,
      \end{equation}
      do not exceed 10\% (i.e. the causal volume of the plasma
      ${\cal
        V}=
      \sigma n_{_{\rm atm}} L \gtrsim 100$), corresponding to an effective cross-sectional area
\begin{equation}
\sigma = \pi r_{_0}^2 = 2.4 \ \times 10^{-10} \
    \left(\frac{M}{{\rm g}} \right)^{2/3} \
    \left(\frac{\rho_s}{\rho_m}\right)^{2/3}~{\rm cm}^2 \gtrsim 3 \times
    10^{-12}~{\rm cm}^2 \, ,
\label{sigma}
  \end{equation}
where $n_{_{\rm atm}}$ is the atmospheric number density and $L \approx c_s t_{\rm cool}$, with
      $c_s \approx 300~{\rm m/s}$ the speed of sound in air and
      $t_{\rm cool}$
      the time required to cool the plasma below  the nitrogen ionization
      temperature $T_{\rm N} \sim 10^4~{\rm
        K}$~\cite{Sidhu:2018auv}. The effective temperature of the
      plasma depends on its radius $r$. In the spirit of~\cite{Cyncynates:2016rij}, we approximate the macro as a delta source and so the initial condition is fixed by equating the heat energy with the macro energy
 \begin{equation}
   T(r,0) = \left|\frac{dE}{dx} \right| \frac{\sigma}{2 \pi \rho_{\rm atm}
     \sigma c_p} \ \frac{\delta(r)}{r} = \frac{\sigma v^2}{2 \pi c_p}
  \ \frac{\delta(r)}{r} \,,
 \end{equation}
where  $c_p \approx 25~{\rm kJ/(kg \, K)}$ is the air specific heat. After
  some time $t$ the temperature field evolves into a Gaussian in $r$
      \begin{equation}
        T (r,t) = \frac{1}{t} \frac{\sigma v}{4 \pi \alpha c_p} \ \exp
        \left[- \frac{r^2}{4 t \alpha} \right] \,,
\label{Trt}
\end{equation}
where $\alpha \approx 10^{-4} \ e^{\zeta/(8~{\rm km})}~{\rm m^2/s}$ is the
  thermal diffusivity of air~\cite{Anchordoqui}.
We can invert (\ref{Trt}) to obtain
  \begin{equation}
\pi r^2 = 4 \pi \alpha t \ln\left(\frac{\sigma \ v^2}{ 4 \pi \alpha t
    c_p T} \right) \,,
\end{equation}
  and then define the cooling time as the time $t>0$ for the
  cylindrical plasma
  to reach zero cross-sectional area at $T = T_N$:
\begin{equation}
t_{\rm cool} = \frac{\sigma v^2}{4 \pi \alpha c_p T_{\rm N}} \, .
\end{equation}
  In
      Fig.~\ref{fig:3} we show characteristic cooling times as a
      function of the cross sectional area. The EUSO-SPB2 fluorescence
      detector integrates over small intervals, known as the bin time,
      $\tau_{\rm bin}$.  For a pixel not to be lit up in multiple
      time bins, we must require $t_{\rm cool} \ll \tau_{\rm bin}$. 

The macro produces a plasma, which persists for a time $t_{\rm cool}$ above about $10^4~{\rm K}$, depositing energy $L \ dE/dx$. As the plasma cools it emits a fraction
 \begin{equation}
        \eta = \frac{2 N_\gamma \overline{E_\gamma}}{\rho_{_{\rm
            atm}} v^2 \sigma L} \sim 2
        \times 10^5 \ \left(\frac{\sigma}{{\rm cm^2} }\right)^2 \
        \left(\frac{v}{250~{\rm km/s}} \right)^4 \ \exp \left(-\frac{ \zeta}{2~{\rm
                km}} \right) \,,
\label{fgamma}
        \end{equation}
            of its energy into photons, where $N_\gamma$ is the number of photons emitted by the plasma and
            $\overline{E_\gamma}$ is the average energy of those
            photons~\cite{Sidhu:2018auv}. From (\ref{uno}) we see that the power dissipation rate of macros going through the
            atmosphere is $\rho_{_{\rm atm}} \sigma v^3$, so the power
            dissipated to UV light  is estimated to be \begin{equation}
  {\cal L} = \eta \ \rho_{_{\rm atm}} \ \sigma \ v^3 = 4.32\times
  10^{-12} \ \left(\frac{M}{\mathrm g}\right)^2 \
  \left(\frac{\rho_s}{\rho_m} \right)^{2}
  \exp\left(-\frac{5\zeta}{8~{\rm km}}\right)~{\rm W}.
\label{L}
\end{equation}  
Using the luminosity (\ref{L}) and requiring that the UV signal
produced by the macro exceeds the noise due to the background photons
at $5\sigma$~\cite{Adams:2017fjh}, a trite
calculation will show that a macro visibility by the EUSO-SPB2
fluorescence detector requires
\begin{equation}
  \sigma \gtrsim 5 \times 10^{-7}~{\rm cm}^2 \, .
\label{sigmalimit}
\end{equation}

 Now, if macros saturate the local DM density, the
expected number of macro events passing through the EUSO-SPB2 FoV is estimated
to be~\cite{Jacobs:2014yca}
\begin{equation}
  {\cal N}  = \frac{\rho_{_{\rm DM}}}{M} \ A \ t_{\rm obs} \ v = 2.4 \
  \left(\frac{M}{3~{\rm g}} \right)^{-1} \  \left(\frac{A}{32~{\rm
        km}^2}\right) \
    \left(\frac{t_{\rm obs}}{360~{\rm hr}} \right) \, ,
\label{NFC}
\end{equation}
where the factor 2.4 reflects the 90\% CL for statistics of small
numbers~\cite{Feldman:1997qc}. Thus, the null result of a EUSO-SPB2
search will yield a 90\%CL
exclusion of a macro flux at $M=3~{\rm g}$ of
\begin{equation}
  {\cal F}_m  < 8.8 \times 10^{-19}~{\rm cm}^{-2} \ {\rm s}^{-1} \ {\rm
    sr}^{-1}  \, ,
\end{equation}
approximating the view of half the sky as $2\pi~{\rm sr}$. This is comparable to the existing upper limit based on indirect
searches~\cite{Price:1988ge,Piotrowski:2020ftp}. The expected 
sensitivity of EUSO-SPB2 for a 100 day flight is  less than  that of
the planned
POEMMA mission after 1 day of operation~\cite{Olinto:2020oky}. A point worth
noting at this juncture is that macros can be easily distinguished
from ordinary meteors. This is because macros travel much faster than
meteors, which being bound to the solar system, travel at
72~km/s relative to the Earth. Furthermore, meteors generally emit
light only in the upper atmosphere where they ablate and
disintegrate. Clear differences in the meteor/macro light
profiles have been observed in numerical
simulations~\cite{Piotrowski:2020ftp,Adams:2014vgr}.

Combining (\ref{sigma}), (\ref{sigmalimit}), and (\ref{NFC})  we can conclude that EUSO-SPB2 will probe
$M \lesssim 3~{\rm g}$ and $\rho_m \lesssim
 10^{-5} \, \rho_s$. Since the sensitivity of EUSO-SPB2 is outside the well-motivated
region of the parameter space ($\rho_m \sim \rho_s$) it is important to
illustrate the boundaries available to $\rho_m \ll \rho_s$. We can envisage that:
\begin{itemize}[noitemsep,topsep=0pt]
\item  the macro is made out of baryons of mass $m_b$;
\item the logarithm of the binding energy per baryon $E_b$ scales
  linearly with the logarithm of the density;
\item the macro density must be well above the atomic density,  $\rho_m \gg  1~{\rm
    g/cm^{3}}$, with $10~{\rm eV} \ll E_b \lesssim 1~{\rm MeV}$.
\end{itemize}
With this in mind, the
macro binding energy scales as $E_b \sim 10~{\rm eV} [\rho_m/({\rm g/cm}^{3})]^{3/7}$,
with $\rho_m  = 3M \pi^{1/2}/ (4 \sigma^{3/2})$. The macro would be
stable while traversing a medium of density $\rho$ and length $l$
provided the total energy transferred be much smaller than the binding energy per baryon multiplied by the number of baryons in the
macro, $E_b M/m_b \gg \rho \sigma v^2 l$. Then, for a macro 
to survive the
passage through the atmosphere we have $\sigma \ll 10^{-4} (M/g)^{20/23}~{\rm cm}^2$. 

In closing, we note that the EUSO-SPB2 fluorescence
detector was designed to measure extensive air showers produced by
ultrahigh-energy cosmic rays. Thereby, the bin time was set at $\tau_{\rm bin} = 1~\mu{\rm s}$, but as shown in Fig.~\ref{fig:3} for the much more slowly moving
macros, we will need to increase this to $\tau_{\rm  bin} \gtrsim 1~{\rm ms}$.

\section{Conclusions}
\label{sec:4}

We have shown that the future EUSO-SPB2 mission has the potential to
detect macroscopic DM. In the case of null result EUSO-SPB2 will
provide a bound comparable and complementary to existing limits.  A
more detailed evaluation of the trigger requirements
($t_{\rm cool} \ll \tau_{\rm bin}$) as well as the potential
implementation is currently under discussion within the Collaboration.

\section*{Acknowledgments}
We thank our colleagues of the EUSO-SPB2 Collaboration for
valuable discussion. TCP and LAA are supported by NASA Grant 80NSSC18K0464. AVO is
supported by NASA Grant 80NSSC18K0246.

\end{document}